\documentclass{kluwer}    
\usepackage{graphics,lscape}
\usepackage[dvips]{graphicx}
\def\Bs{\char92}

\usepackage{makeidx}

\makeindex

\newcommand{\Eindex}[1]{\index{Environments!#1}\index{#1 environment}}

\newcommand{\Cindex}[1]{\index{#1@\texttt{\Bs #1}}}

\begin{document}                                                                                   
\begin{article}
\begin{opening}         
\title{Small-Angle Scattering and Diffusion: Application to Relativistic Shock Acceleration}
\author{R.J. \surname{Protheroe}, A. {Meli}\thanks{visiting from Imperial College, London} and A.-C. \surname{Donea}}  
\runningauthor{R.J.\ Protheroe, A.\ Meli \& A.-C.\ Donea}
\runningtitle{Small-Angle Scattering and Diffusion: Application to Relativistic Shock Acceleration}
\institute{Department of Physics \& Mathematical Physics, The University of Adelaide, Adelaide, SA 5005, Australia}
\date{19 June 2002}

\begin{abstract}
We investigate ways of accurately simulating the propagation of
energetic charged particles over small times 
where the standard Monte Carlo approximation to diffusive
transport breaks down.  We find that a small-angle
scattering procedure with appropriately chosen step-lengths and
scattering angles gives accurate results, and we apply this to
the simulation of propagation upstream in relativistic shock
acceleration.
\end{abstract}

\end{opening}           

\section{Introduction}  

Relativistic charged particle transport in magnetized
astro-physical plasma is strongly affected by magnetic
irregularities, and may be approximated by diffusion.  Diffusive
transport of particles having speed $v$ can be simulated by a
three-dimensional random walk with steps sampled from an
exponential distribution with mean free path $\lambda=3D/v$,
where $D$ (cm$^2$ s$^{-1}$) is the spatial diffusion coefficient,
followed by large-angle (isotropic) scattering after each step
(e.g.\ Chandrasekhar 1943), and this gives good results for
distances much larger than $\lambda$.

In diffusive shock acceleration at relativistic shocks problems
arise when simulating particle motion upstream of the shock
because the particle speeds, $v$, and the shock speed $v_{\rm
shock}=c(1-1/\gamma_{\rm shock}^2)^{1/2}$ are both close to $c$,
and so very small deflections are sufficient to cause a particle
to re-cross the shock.  Clearly, Monte Carlo simulation by a
random walk with mean free path $\lambda$ and large-angle
scattering is inappropriate here, and in Monte Carlo simulations
of relativistic shock acceleration at parallel shocks Achterberg
et al.\ (2001) consider instead the diffusion of a particle's
direction for a given angular diffusion coefficient $D_\theta$
(rad$^2$ s$^{-1}$).  Similarly, for a given spatial diffusion
coefficient $D$, Protheroe (2001) and Meli \& Quenby (2001) adopt
a random walk with a smaller mean free path, $\bar{\ell} \ll
\lambda$, followed by scattering at each step by a small angle
with mean deflection , $\bar{\theta} < 1/\gamma_{\rm shock}$.
See Bednarz \& Ostrowski (2001) for a recent review of
relativistic shock acceleration.

\section{Small-angle scattering and diffusion}

We consider propagation by small steps sampled from an
exponential distribution with mean $\bar{\ell} \ll \lambda$, followed at each
step by scattering by a small angle sampled from an
exponential distribution with mean $\bar{\theta} \ll \pi$.
The change in direction ($\theta_1,\theta_2$) may then be described
as two-dimensional diffusion with angular diffusion coefficient
$D_\theta=\bar{\theta}v_\theta/2$ (rad$^2$ s$^{-1}$) where
$v_\theta=\bar{\theta}/\bar{t}$, and $\bar{t}=\bar{\ell}/v$ such
that $D_\theta=\bar{\theta}^2v/(2\bar{\ell})$.  The time $t_{\rm
iso}$, which gives rise to a deflection equivalent to a
large-angle (isotropic) scattering, is determined by
$(\sigma_{\theta_1}^2+\sigma_{\theta_2}^2)^{1/2}=\sqrt{4D_\theta t_{\rm iso}} \sim \pi/2$, giving $\lambda \sim
vt_{\rm iso} \propto \bar{\ell}/\bar{\theta}^2$ and  a spatial
diffusion coefficient $D \propto \bar{\ell}v/\bar{\theta}^2 \propto
v^2/D_\theta$.
By using a Monte Carlo method it is straightforward to test this,
determine the constant of proportionality, and thereby make the
connection between diffusion and small angle scattering.  

\begin{figure}[t]
\centerline{\includegraphics[width=4.5in]{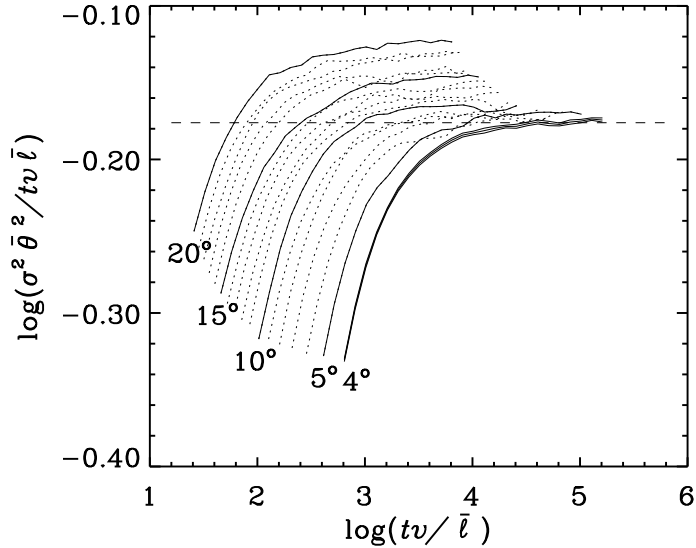}}
\caption{$\sigma^2$ vs.\ time for a 3D random walk with isotropic
injection at the origin at $t=0$. Step-lengths $\ell$ were
sampled from an exponential distribution with mean $\bar{\ell}$
followed by small-angle scattering with scattering angle $\theta$
sampled from an exponential distribution with mean $\bar{\theta}$
(the numbers attached to the curves).  Dashed line is $\sigma ^2=
2 t v \bar{\ell} /3 \bar{\theta}^2 $.  Curves for
5$^\circ$--20$^\circ$ result from $10^4$ simulations; 4$^\circ$
curve results from $8\times 10^4$ simulations (width shows
statistical error). }\label{fig1}
\end{figure}

The solution of the diffusion equation for a delta-function
source in position and time
$q(\vec{r},t)=\delta(\vec{r})\delta(t)$ and an infinite diffusive
medium is a three-dimensional gaussian with standard deviation
$\sigma=\sqrt{2Dt}$ (Chandrasekhar 1943).  The results from
several Monte Carlo random walk simulations are shown in
Fig.~\ref{fig1}, from which we find that the expected dependence
occurs for $\bar{\theta} < 5^\circ$ at times $t > 10^5
\bar{\ell}/v$.  For this case, we see from Fig.~\ref{fig1} that
$\sigma^2 \to 2 t v \bar{\ell} /3 \bar{\theta}^2 $, and so we
obtain the connection between small-angle scattering and
diffusion theory, namely $D \approx \bar{\ell}v/(3\bar{\theta}^2)
\approx v^2/(6D_\theta)$.

\Eindex{figure}\Cindex{caption}\Cindex{centerline}

\vspace*{-0.5em}

\section{Application to relativistic shock acceleration}

As viewed in the frame of reference of the upstream plasma,
ultra-relativistic particles are only able to cross the shock
from downstream to upstream if the angle $\theta$ between their
direction and the shock normal pointing upstream is $\theta <
\sin^{-1}(1/\gamma_{\rm shock})$, where $\gamma_{\rm
shock}=(1-\beta_{\rm shock}^2)^{-1/2}$ and $\beta_{\rm
shock}=v_{\rm shock}/c$.  For highly relativistic shocks these
particles cross the shock from downstream to upstream travelling
almost parallel to the shock normal.  Similarly, having crossed
the shock, only a very slight angular deflection, by $\sim
1/\gamma_{\rm shock}$ is sufficient to return them downstream of
the shock.  This change in particle direction gives rise to a
change in particle energy $E'$ and momentum $p'$, measured in the downstream
plasma frame (primed coordinates), of
\begin{eqnarray}
{E_{n+1}' \over E_{n}'} \approx {p_{n+1}' \over p_{n}'} =
{1-\beta_{\rm 12} \cos\theta_{n+1} \over 1-\beta_{\rm 12} \cos\theta_{n}}
\end{eqnarray}
in an acceleration cycle (downstream $\to$ upstream $\to$
downstream), where $\beta_{\rm 12}$ is the speed of the upstream
plasma as viewed from the downstream frame.

\begin{figure}[H]
\tabcapfont
\centerline{%
\begin{tabular}{c@{\hspace{-5.5pc}}c}
\includegraphics[width=3.4in]{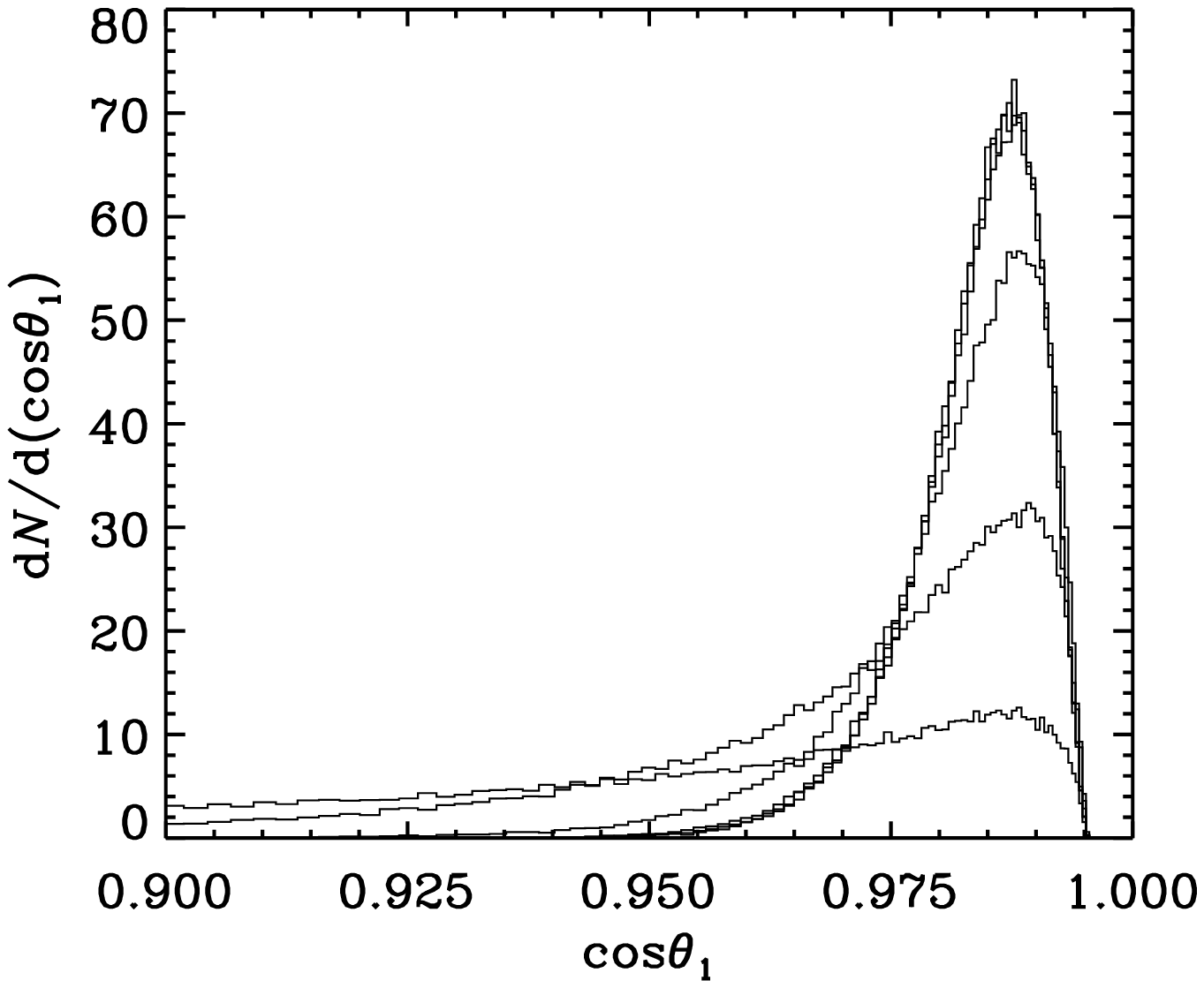} &
\includegraphics[width=3.4in]{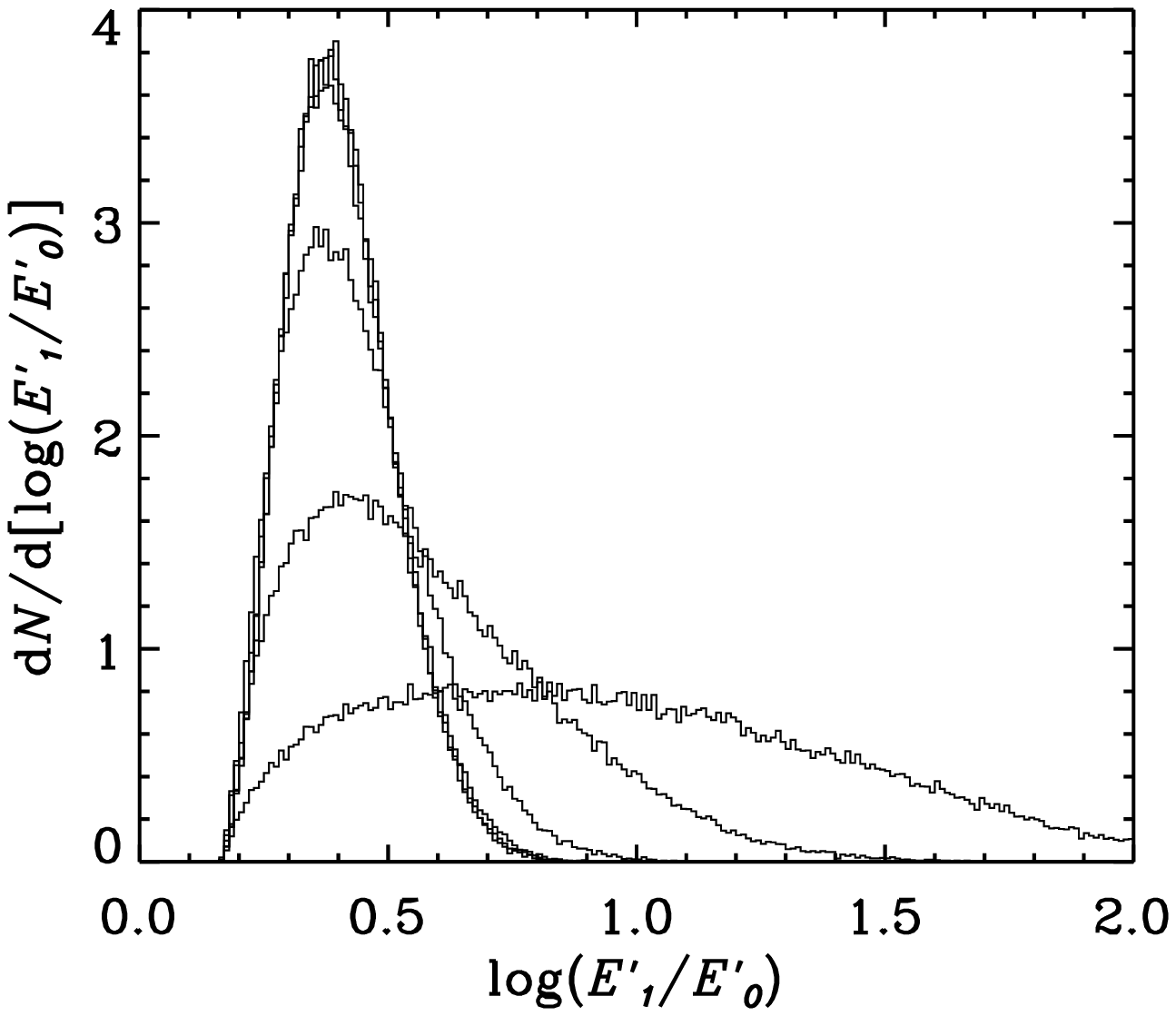} \\
a.~~ $\cos\theta_1$ distribution & b.~~ $\log(E'_1/E'_0)$ distribution
\end{tabular}}
\caption{Small-angle scattering simulation of excursion upstream
in diffusive shock acceleration at a parallel relativistic shock
with $\gamma_{\rm shock}=10$.  Results are shown for $10^5$ injected particles and
$\bar{\theta}=10^{-2}/\gamma_{\rm shock}$ (top histogram),
$3\times 10^{-2}/\gamma_{\rm shock}$, $10^{-1}/\gamma_{\rm
shock}$, $0.3/\gamma_{\rm shock}$, $1/\gamma_{\rm shock}$ and $3/\gamma_{\rm shock}$
(bottom histogram).  Note that the top three histograms are almost indistinguishable.}
\label{fig2}
\end{figure}

In ``parallel shocks'' the magnetic field is parallel to the
shock normal, and so the the pitch angle $\psi$ is the angle to
the shock normal and $v\cos\psi$ gives the component of velocity
parallel to the shock.  Thus the small-angle scattering method
described in the previous section is used here to simulate
particle motion upstream of a parallel relativistic shock,
including the effects of pitch-angle scattering, for a given
diffusion coefficient.  We inject ultra-relativistic particles at
the shock with downstream-frame energy $E'_0$ travelling upstream
parallel to the shock normal, i.e.\ $\theta_0 \! = \! 0$.  We
follow a particle's trajectory until the shock catches up with
it, and it crosses from upstream to downstream with an
upstream-frame angle $\theta_1$ to the shock normal and a
downstream-frame energy $E'_1$.  The simulation was performed for
$\gamma_{\rm shock}=10$, and five different mean scattering
angles $\bar{\theta}$, to determine the maximum
$\bar{\theta}$-value that can safely be used for accurate
simulation.  The resulting distributions of $\cos\theta$ and
$\log(E'_1/E'_0)$ are shown in Fig.~\ref{fig2}, and show that in
this application one requires $\bar{\theta} < 0.1/\gamma_{\rm
shock}$.  Our results are quite consistent with those of
Achterberg et al.\ (2001), who used a diffusive angular step
$\Delta \theta_{\rm st} \le 0.1/\gamma_{\rm shock}$.

\vspace*{-2em}
\section{Conclusion}

The standard Monte Carlo random walk approach to simulation of
energetic charged particle propagation for a given spatial
diffusion coefficient $D$ can be extended to apply accurately to
times much less than $\lambda/v = 3D/v^2$ by using a small-angle
scattering procedure with steps sampled from an exponential
distribution with mean free path
$\bar{\ell}=\bar{\theta}^2\lambda$ followed at each step by
scattering with angular steps sampled from an exponential
distribution with mean scattering angle $\bar{\theta} < 0.09$~rad
($5^\circ$).  The spatial and angular diffusion coefficients are
then $D \approx \bar{\ell}v/(3\bar{\theta}^2)$ and
$D_\theta=\bar{\theta}^2v/(2\bar{\ell})$, and are related by $D
\approx v^2/(6D_\theta)$.  In simulation of upstream propagation
in relativistic shock acceleration one must use $\bar{\theta} <
0.1/\gamma_{\rm shock}$ to obtain accurate results.

\end{article}
\end{document}